\documentclass[conference, a4 paper]{IEEEtran}
\ifCLASSINFOpdf
   \usepackage[pdftex]{graphicx}
\else
   \usepackage[dvips]{graphicx}
\fi
\usepackage[cmex10]{amsmath}
\usepackage{gensymb}

\makeatletter
\usepackage{lipsum}
\usepackage[utf8]{inputenc}
\usepackage{longtable}
\usepackage{comment}

\usepackage{textcomp}
\usepackage{float}
\usepackage{mathtools}
\graphicspath{ {./Images/} }
\usepackage{blindtext}
\def\BibTeX{{\rm B\kern-.05em{\sc i\kern-.025em b}\kern-.08em T\kern-.1667em\lower.7ex\hbox{E}\kern-.125emX}}
\usepackage{enumitem}
\usepackage{subcaption}
\usepackage[
backend=biber,
style=ieee,
sorting=none
]{biblatex}
\begin{document}
\setlength{\parskip}{0pt}
 \setlength{\textfloatsep}{1pt}
 \setlength{\abovecaptionskip}{2pt} 
 \setlength{\belowcaptionskip}{2pt} 
%

\title{Exhaustive Classification and Quantification of Coupling Modes in Power Systems with Power Electronics}

\author{\IEEEauthorblockN{Pamela Zoghby}
	\IEEEauthorblockA{\textit{Électricité De France (EDF R\&D)} \\
\IEEEauthorblockA{{Palaiseau, France}} 
\textit{Ecole Centrale Nantes,} \\
    \textit{CNRS, LS2N UMR 6004, Nantes}\\
		pamela.zoghby@edf.fr\\}
	\and
	\IEEEauthorblockN{Bogdan Marinescu}
	\IEEEauthorblockA{\textit{Nantes Université} \\
    \textit{Ecole Centrale Nantes}\\
    \textit{CNRS, LS2N UMR 6004}\\
		Nantes, France \\
		bogdan.marinescu@ec-nantes.fr}
  	\and
	\IEEEauthorblockN{Antoine Rossé, Grégoire Prime}
	\IEEEauthorblockA{\textit{Électricité De France (EDF R\&D)} \\
		Palaiseau, France \\
		antoine.rosse@edf.fr \\}
}


\maketitle

\begin{abstract}
Due to the energy transition, today's electrical networks include synchronous machines and inverter-based resources interfacing renewable energies such as wind turbines, solar panels, and Battery Energy Storage Systems to the grid. In such systems, interactions known as coupling modes or dynamic interactions, between synchronous machines and inverter-based resources may arise. This paper conducts a clear and exhaustive study on a proposed benchmark, in order to analyze, quantify and classify these new types of modes. Detailed models representing electromagnetic transient phenomena are developed and linearized, then used for conducting modal analysis to fully characterize the small-signal stability of the system. Also, a sensitivity analysis is presented to evaluate the impact of key parameters on the detected modes of oscillation. Besides the exhaustive classification of the possible coupling modes, the proposed benchmark and methodology can be used to study any given power system in a minimal order modeling. The case of a fully detailed power grid based on the IEEE 39-bus system was studied as an illustrative example.
\end{abstract}



\section{Introduction}
Because of worldwide climate change concerns, conventional fossil-fuel power plants based on synchronous machines (SMs) are gradually being replaced by renewable energies such as solar photovoltaic (PV) and wind turbines, as well as battery energy storage systems (BESS), all of which are inverter-based ressources (IBRs).

The integration of these technologies introduces challenges related to dynamic interactions, often referred to as \textit{coupling modes}. These coupling modes arise due to the contrasting characteristics of IBRs and SM. 
Indeed, IBRs include high-frequency switching, short response times, in addition to slower dynamic ranges. Some of these dynamics resemble those of SM, potentially leading to interactions that can affect the stability and reliability of the power system. 

Similarly to the inter-area modes, coupling modes can cause wear and tear on a power system's components and may trigger stability issues, potentially leading to power plant tripping and even black-outs in extreme cases, if not well damped.

 As far as the authors are aware, no prior research has delved into a comprehensive and exhaustive interpretation of these coupling modes. Prior studies have either focused on the stability of 100 \% inverter-based systems \cite{Peng2019,huang2018,Gautam2009}, or have studied only the stability margins of the system with a certain penetration of IBRs \cite{Markovic2021}. Some authors explored systems comprising a SM and an IBR \cite{Pattabiraman2018a,Lin2017,Pereira2019,Ding2021}, however the case studies employed are simplified and not representative of a real power system. 

Authors in \cite{Pattabiraman2018a} and \cite{Lin2017}
analyse an oversimplified system comprised of a SM and an inverter. Stator, networks transients and current controller dynamics of the grid-following (GFL) inverter are ignored in \cite{Pattabiraman2018a}, and stator
damper windings are disregarded in \cite{Lin2017}. It is possible that due to model over-simplification, certain dynamic coupling modes may be missed in the analyses. 

Reference \cite{Pereira2019} shows only the critical eigenvalues having a damping less that 5\% of a system comprised of a single inverter connected to a single SM. In \cite{Ding2021}
authors were interested in dynamic couplings between SM and inverters. However, both references \cite{Pereira2019} and  \cite{Ding2021} relied on simplified case studies that, by disregarding the effect of the rest of the power system, may lead to inaccurate or incomplete findings.

This paper aims to thoroughly analyze, classify, and understand the behavior of the coupling modes using a proposed exhaustive benchmark system representative of a real power system. The benchmark is composed of a SM, an IBR and an equivalent grid to model the rest of the system. The proposed benchmark system is sufficient to analyse the dynamic interactions and coupling modes yet minimal to reduce complexities. The study is exhaustive as the benchmark is designed to capture a wide variety of scenarios that depict different types of modes.

Also, a sensitivity analysis is conducted to evaluate the influence of various system parameters on these modes. Equipped with this information, system operators 
will possess an understanding of these modes, the factors impacting them and can implement measures to prevent system instability caused by their low damping. Furthermore, power system engineers may benefit from this benchmark as a practical minimal model to which any real and complex power system can be reduced for the initial study of the coupling modes, as demonstrated in its application to the IEEE 39-bus case study discussed in section V.  

The remaining sections of the paper are organised as follows: section II describes the utility of the benchmark and its structure. Section III shows the coupling modes obtained by performing modal analysis. In section IV, the coupling modes are classified as structural and non-structural according to their sensitivity to main system parameters. In section V, the coupling modes of a modified IEEE 39-bus system are detected and analyzed using the benchmark. Section VI is devoted to conclusions.

\section{The Benchmark and its Structure}
\begin{figure}[h]
\centering
 \captionsetup{justification=centering}
\begin{subfigure}[b]{0.45\textwidth}
\centering
\includegraphics[width=8cm]{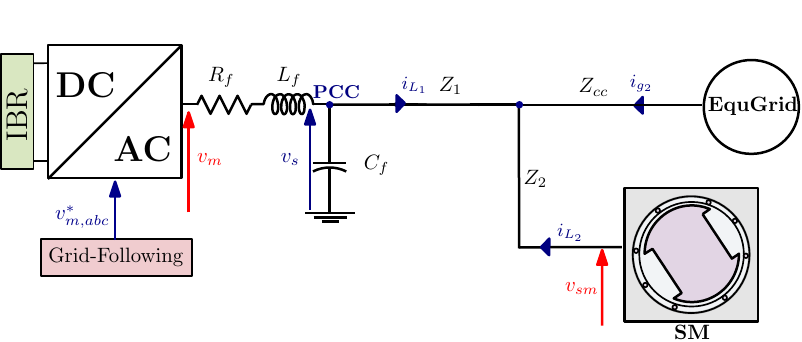}
\caption{Star form}
\label{fig:etoile}  
\end{subfigure}
\begin{subfigure}[b]{0.45\textwidth}
\centering
\includegraphics[width=8cm]{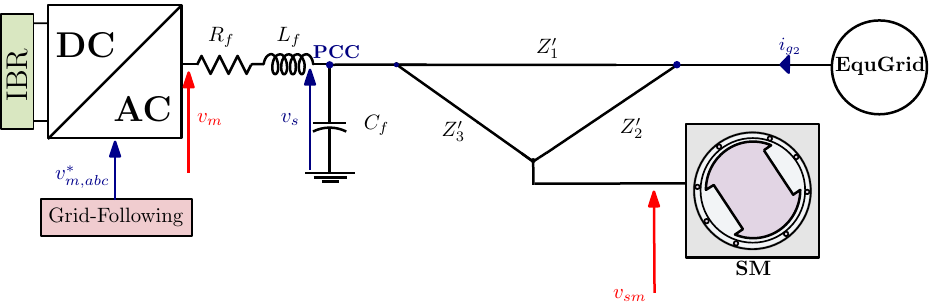}
\caption{Triangle form}
\label{fig:triangle}  
\end{subfigure} 
 \caption{\footnotesize{A benchmark representing a real power system.}}
\label{image1}
\end{figure}

An exhaustive but minimal benchmark is proposed to fully study, analyse and quantify coupling modes between traditional SMs and IBRs. When considering only one SM and one IBR within a power grid, the structure is illustrated in Fig.\ref{image1}.

First, the electrical distances are captured by representing the equivalent impedances between the different components, considering thus their potential impact on the coupling modes. As a consequence, the benchmark's configuration is in the form of a star as in Fig. \ref{image1}(\subref{fig:etoile}) or a triangle as in Fig. \ref{image1}(\subref{fig:triangle}). Both structures are equivalent and, for example, switching from (a) to (b) is done using equations (\ref{eqn 17}).
\begin{equation}
	\begin{aligned}
		 {Z'_1}=\frac{Z_1Z_{cc} + Z_2Z_{cc} + Z_1Z_2}{Z_2}\\
      {Z'_2}=\frac{Z_1Z_{cc} + Z_2Z_{cc} + Z_1Z_2}{Z_1}\\
      {Z'_3}=\frac{Z_1Z_{cc} + Z_2Z_{cc} + Z_1Z_2}{Z_{cc}}\\
	\end{aligned} 
 	\label{eqn 17}
\end{equation} 
Using one or the other depends on the situation to be studied. For example, if the impact of the distance between the SM and IBR is to be studied, configuration in Fig.\ref{image1}(\subref{fig:triangle}) is used. However, configuration in Fig.\ref{image1}(\subref{fig:etoile}) is used if the distance to the rest of the system is of interest. The electrical distances between the different elements are named as follows:
\begin{itemize}
  \item $Z_1'$ the electrical distance between the inverter and the equivalent grid.
  \item $Z_2'$ the electrical distance between the SM and the equivalent grid.
  \item $Z_3'$ the electrical distance between the inverter and the SM.
  \item $Z_{cc}$ the electrical distance between equivalent grid and the other elements of the benchmark.
\end{itemize}

This structure is inherited from \cite{Marinescu2021} where a new \textit{control model} is proposed to synthesize new controls for IBRs.

Next, to capture the influence of voltage and frequency dynamics of the grid within the benchmark, a representation of the rest of the system is necessary. It is worth noting that in
systems with high shares of IBRs, the voltage and frequency
dynamics are not necessarily clearly separated as is generally
the case in traditional power systems dominated by SMs. Thus the block EquGrid of the benchmark is an original grid equivalent model representing the rest of the system which includes the loads and other generation units, whether they be synchronous generators or IBRs. Compared to classic representations (e.g., infinite bus) this equivalent accounts for both voltage and frequency dynamics \cite{Marinescu2021}.

The proposed benchmark structure thus improves the simplified models that do not include any model of the rest of the system used in previous works \cite{Peng2019,huang2018,Gautam2009}. As demonstrated further in this paper, it enabled the authors to comprehensively classify the coupling modes and quantify their sensitivities to different grid situations and configurations.

Furthermore, in order to preserve all dynamics in the study, all the components of the benchmark are modeled in EMT (Electromagnetic Transient). 

To be exhaustive, all the benchmark parameters are varied in sufficiently large intervals. These variations are performed starting from a nominal situation given in Table \ref{tab:table3} of Appendix A. As the line impedances $Z_i$ are defined in Ohm/km, $l_i$ represents the length of the line between the different elements, measured in km.

As this benchmark is exhaustive, it can also be used to simplify the model of a given power system in which a given set of IBRs are to be inserted. This provides a way to run a simplified preliminary study of the coupling modes as demonstrated in Section V for the IEEE 39-bus power system.



\subsection{Synchronous machine's model and its regulations}
A conventional SM with its voltage and frequency regulations is used. The detailed model of the SM with the dynamics of the stator and dampers is adopted and all the electrical equations are provided in \cite{Markovic2021}, \cite{Ding2021} and \cite{Pereira2020a}. 

The mechanical dynamics of the SM are described by the swing equations in \eqref{eqn 7} and \eqref{eqn 8}.
\begin{equation}
	\label{eqn 7}
	\begin{aligned}
		\frac{d\omega_{SM}}{dt}&= \frac{1}{2H}(P_m-P_e-{K_d}\Delta \omega_{SM})
	\end{aligned}
\end{equation}
\begin{equation}
	\label{eqn 8}
	\begin{aligned}
		\frac{d\theta_{SM}}{dt}&= \omega_b\Delta \omega_{SM}
	\end{aligned}
\end{equation}
where $H$ and $K_d$ are the inertia constant and damping coefficient of the SM respectively. $P_m$ and $P_e$ are the mechanical and electrical power. $\Delta \omega_{SM}$ is the rotor frequency deviation with respect to the grid frequency, and $\theta_{SM}$ is the internal angle of the SM.

\begin{figure}[h]
\centerline{\includegraphics[width=8cm]{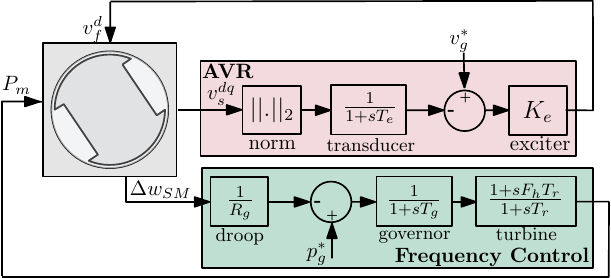}}
 \captionsetup{justification=centering}
		\caption{\footnotesize{SM's AVR and frequency control.}}
		\label{image2}
\end{figure}

The voltage is regulated by a static-type Automatic Voltage Regulator (AVR) simplified to the first order as shown in Fig. \ref{image2}
with $T_{e}$ the transducer time constant, ${v_g}^*$ the voltage setpoint, and ${v_f}$ the regulated field voltage.

The frequency control consists of a droop, a governor and a turbine also simplified to the first order as shown in Fig. \ref{image2}.
\subsection{Grid-following inverter's model}
\begin{figure}[h]
\centerline{\includegraphics[width=8cm]{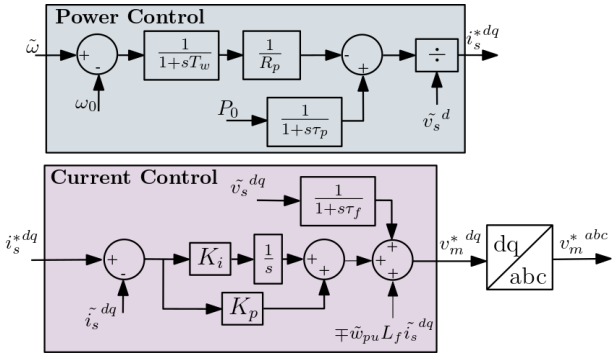}}
 \captionsetup{justification=centering}
		\caption{\footnotesize{Grid-Following Control block diagram}}
		\label{image3}
\end{figure}

 For this article, only the grid-following control mode is considered, as it still remains the predominant control type of the majority of IBRs in today's power systems, despite growing interest for grid-forming control that is held for future works. As illustrated in Fig. \ref{image3}, the grid-following control includes an inner current loop modeled as a PI controller, with $K_i$ and $K_p$ the integral and proportional gains respectively. A low-pass filter with time constant $\tau_f$ is added to the voltage feed-forward to increase the robustness of the inverter as proposed in \cite{Pereira2020a}.

An outer power control loop gives the current references for the current controller. A low-pass filter of time constant $\tau_p$ represents the power setpoint dynamics. A power-frequency droop is added to allow the conveter to participate in primary frequency control. The low-pass filter with time constant $\tau_w$ reduces frequency oscillations that may be caused by the inverter's Droop dynamics \cite{Pereira2020a}.

A PLL (Phase Locked Loop) is used for synchronisation of the inverter with the grid and to estimate the grid frequency. The PLL is composed of a PI controller driving the component ${\tilde{v_s}}^{q}$ to zero, thus creating a reference frame (dq), aligned with $\overrightarrow{v_s}$ as in \cite{Pereira2020a}.
\subsection{Lines and load model}
The lines are modelled as a resistance in series with an inductance taking into consideration all EMT dynamics and represented in dq frame  as in \eqref{eqn 11}.
\begin{equation}
	\label{eqn 11}
	\begin{aligned}
		\frac{di_k^{dq}}{dt}&= v_i^{dq} - v_j^{dq} - R_ki^{dq} \pm L_ki^{qd}
	\end{aligned}
\end{equation}
\subsection{Equivalent grid model}
The rest of the system is represented by the block EquGrid. As stated in \cite{Marinescu2021}, the latter is composed of a three-phase voltage source \eqref{GDEvoltage} and a swing equation \eqref{eqn 12} that forms the grid frequency, where $H_e$ and $D_u$ are the global inertia and the global damping coefficient of the remainder of the power system respectively.  

This combination represents the aggregate dynamic behavior of the rest of the network.
\begin{equation}\label{GDEvoltage}
\begin{array}{l}
\begin{aligned}
	{V_B}^a=V\sin(\theta_f)\\
	{V_B}^b=V\sin(\theta_f-2\frac{\pi}{3})\\
	{V_B}^c=V\sin(\theta_f+2\frac{\pi}{3})
\end{aligned}
	\end{array}
\end{equation}

\begin{equation}
	\label{eqn 12}
	\begin{aligned}
		\frac{d\omega_f}{dt}&= \frac{1}{2H_e}(P_{G}-P_l-D_u\Delta \omega_f), \quad   \omega_f= \frac{d\theta}{dt}\\
	\end{aligned}
\end{equation}
\\

\section{Modal analysis}
Joint nonlinear time-domain simulations and modal analysis have been performed for investigation. Indeed, nonlinear time-domain simulations cannot fully detect highly damped oscillation modes, and provide limited information about the type of the observed modes. To address these limitations, modal analysis is used, requiring the linearisation of some inherently nonlinear equations of the system developed in section II. Conversely, all results obtained via modal analysis are verified by well chosen nonlinear time-domain simulations.

The equations of the entire system are put in the form of differential-algebraic equations as in \eqref{eqn 13}. 
\begin{equation}
	\label{eqn 13}
	\begin{aligned}
		 \dot{x}&=f(x,u) \quad
           0=g(x,u)\\
	\end{aligned}
\end{equation}
They are next linearised around an operating point obtained from the load flow and put into state space form as in \eqref{eqn 14}.
\begin{equation}
	\label{eqn 14}
	\begin{aligned}
		 \dot{x}&=Ax + Bu\\
	\end{aligned}
\end{equation}

Using matrix A, modal analysis enables the determination of the frequency and damping of oscillation modes, identification of the mode type by assessing participation of the states, sensitivity analysis to identify the parameters affecting the mode and an extended mode shape analysis to assess how each mode affects specific state variables.

An extended mode shape analysis was introduced in \cite{miao2023} to quantify modal contributions of electric variables which have components in both \textit{d} and \textit{q} axes. For such a variable $\Gamma_m$, for example the magnitude of a current, its mode shape is 

\begin{equation}
	\label{eqn 30}
	\begin{aligned}
		 \frac{\partial I_m}{\partial I_d}\Gamma_d + \frac{\partial I_m}{\partial I_q} \Gamma_q 
	\end{aligned}
\end{equation}

where $\Gamma_d$ and $\Gamma_q$ are the right eigenvectors of the \textit{d} and \textit{q} axis current respectively.

The linear model (\ref{eqn 14}) is initially established for a specific scenario corresponding to $l_{1}$ = $l_{2}$ = 5 km and $l_{cc}$ = 20 km, with the electrical impedance parameters provided in Appendix A, Table \ref{tab:table4}. In the next section, several variations of this situation are considered. All the known classical modes of oscillation (e.g. electromechanical mode, frequency mode, mode local to the SM) have been identified and largely studied in other works \cite{Kundur2004,Rogers2015,Liu2017}, therefore, only the coupling modes
between the SM and the inverter are of interest in this paper as they require further analysis, and they are displayed in Table~\ref{tab:table1}. 
\begin{table}[h]
\begin{center}
\captionsetup{justification=centering}
\caption{\\ \small{Relevant modes obtained by the modal analysis approach}}
\addtolength{\tabcolsep}{-1pt} 
\begin{tabular}{ | c | c| c | c |} 
  \hline
\textbf{Mode} & \boldmath{$\lambda = \sigma \pm j\omega$} & \textbf{Frequency (Hz)} & \textbf{Damping (\%)}\\ 
  \hline
  1 & $-3082 \pm 1395j$  & 222 & 26  \\ 
  \hline
  2 & $-632.57 \pm 1450j$  & 230 & 44  \\ 
  \hline
  3 & $-1433 \pm 99j$ & 16 & 99  \\ 
  \hline
  4 & $-1345 \pm 3927j$ & 625 & 32  \\ 
  \hline
  5 & $-1712 \pm 3462j$ & 551 & 44  \\ 
  \hline
		\end{tabular}
\label{tab:table1}
	\end{center}
\end{table}

As shown in Table \ref{tab:table1}, these modes exhibit high frequencies compared to typical electromechanical modes and demonstrate strong damping characteristics under the initial scenario. 

\begin{figure*}[!h]
\centering
 \captionsetup{justification=centering}

\begin{subfigure}[b]{0.35\textwidth}
\centering
\includegraphics[width=\textwidth]{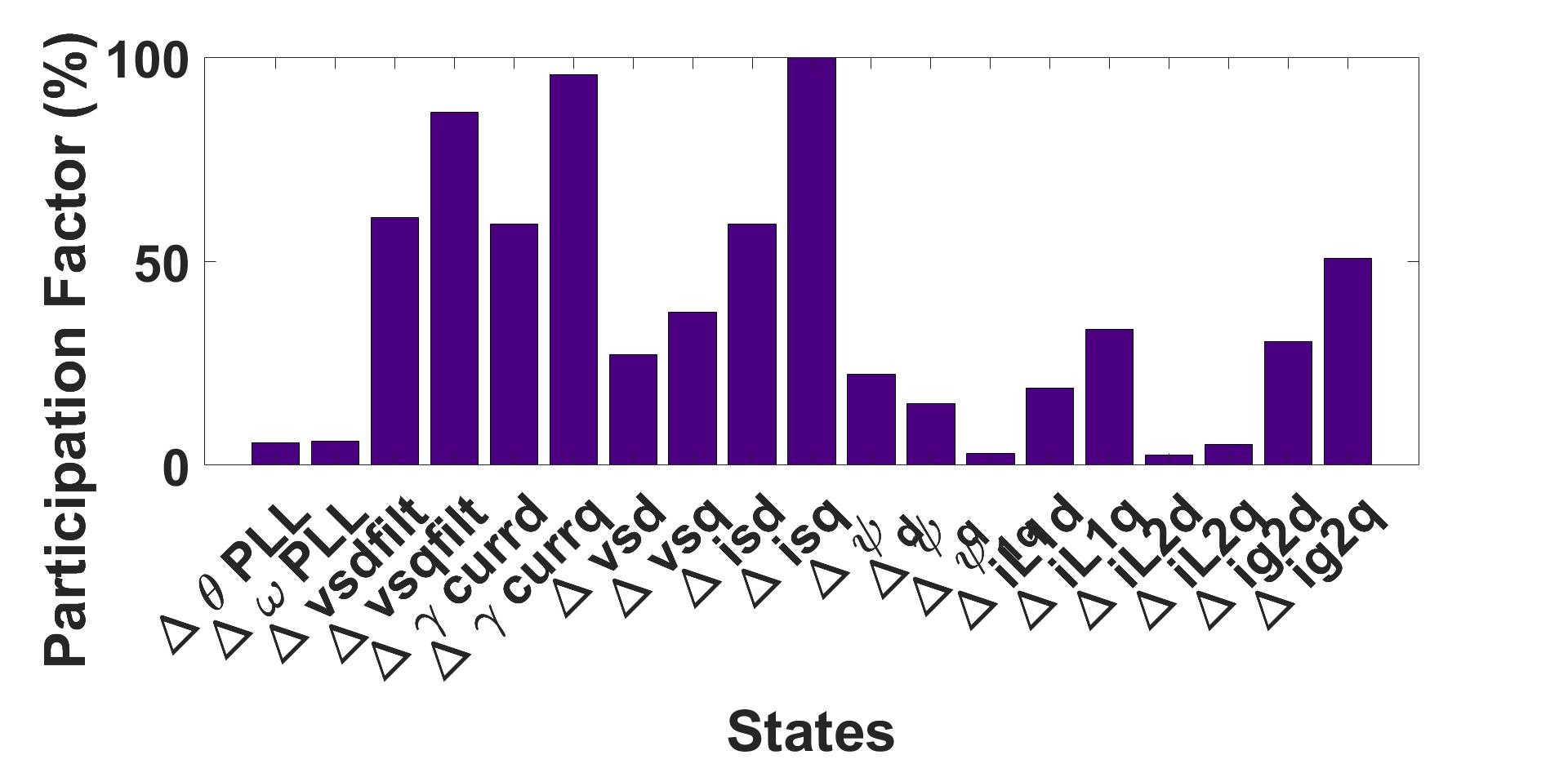}
\caption{Mode 1}
\label{fig:mode 2}  
\end{subfigure}    
\hspace{-0.8cm}
\begin{subfigure}[b]{0.35\textwidth}
\centering
\includegraphics[width=\textwidth]{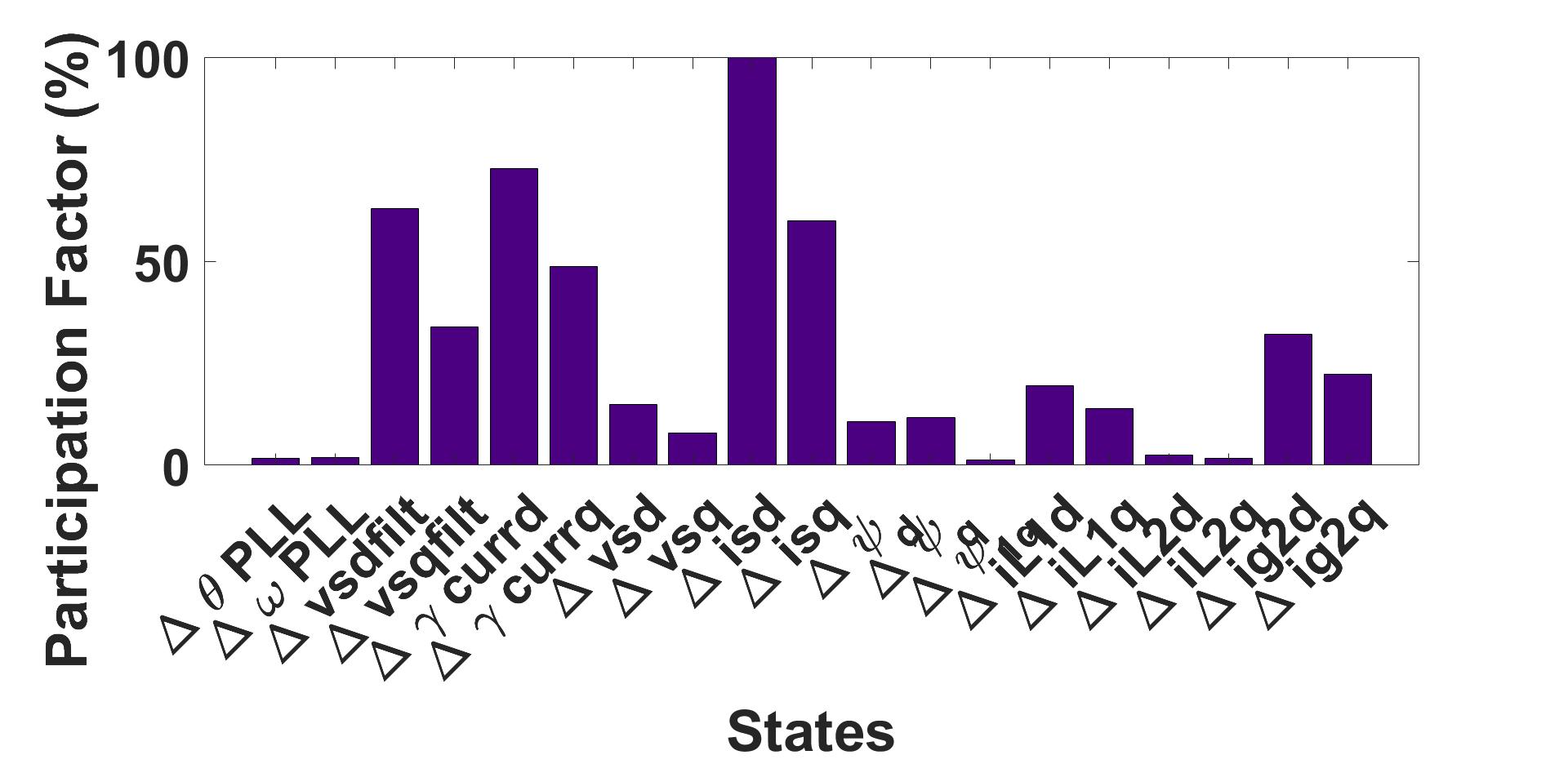}
\caption{Mode 2}
\label{fig:mode 3}  
\end{subfigure}
\hspace{-0.8cm}
\begin{subfigure}[b]{0.35\textwidth}
\centering
\includegraphics[width=\textwidth]{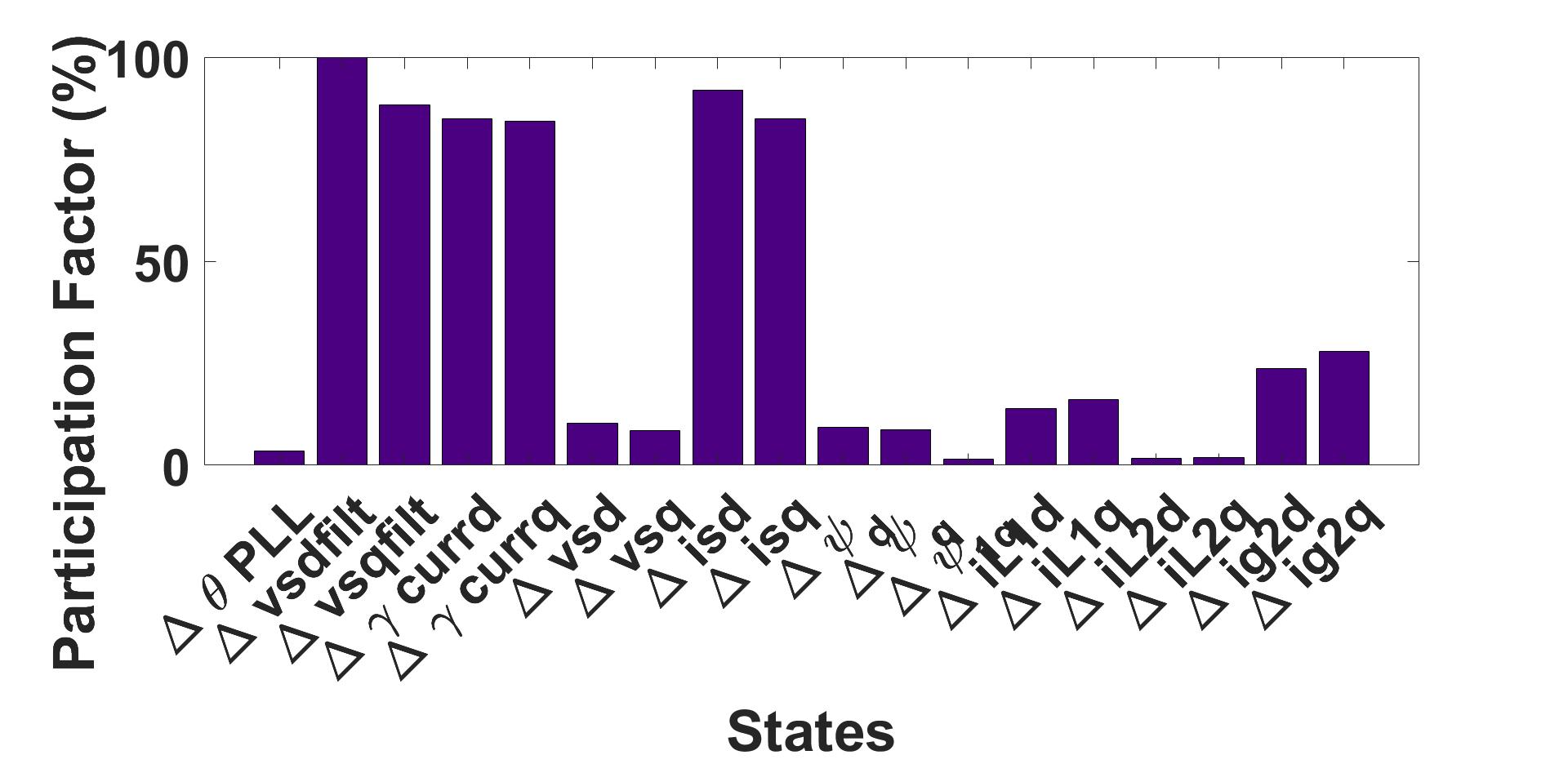}
\caption{Mode 3}
\label{fig:mode 4}  
\end{subfigure}
\\
\begin{subfigure}[b]{0.35\textwidth}
\centering
\includegraphics[width=\textwidth]{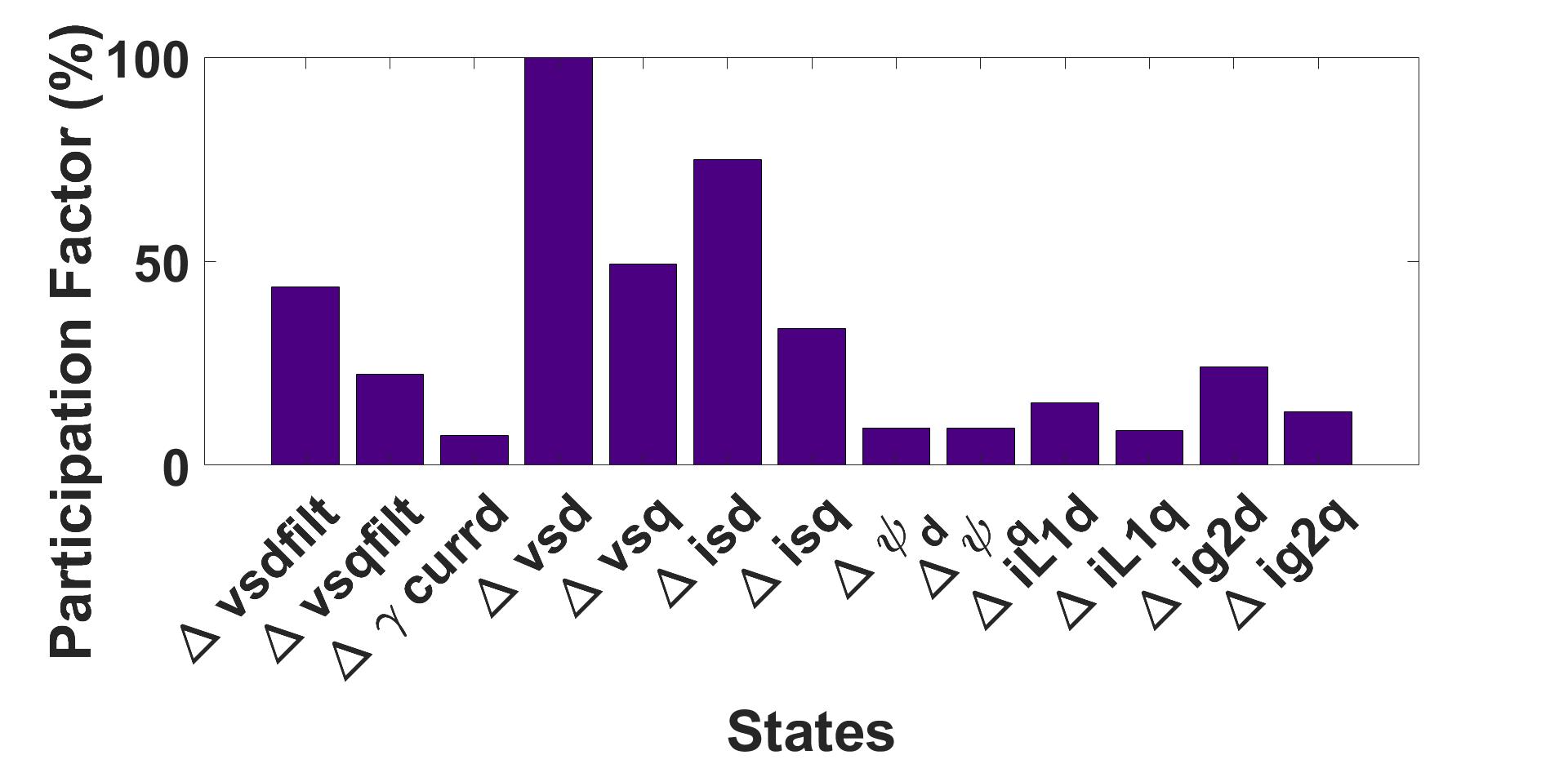}
\caption{Mode 4}
\label{fig:mode 5}  
\end{subfigure}
\hspace{-0.8cm}
\begin{subfigure}[b]{0.35\textwidth}
\centering
\includegraphics[width=\textwidth]{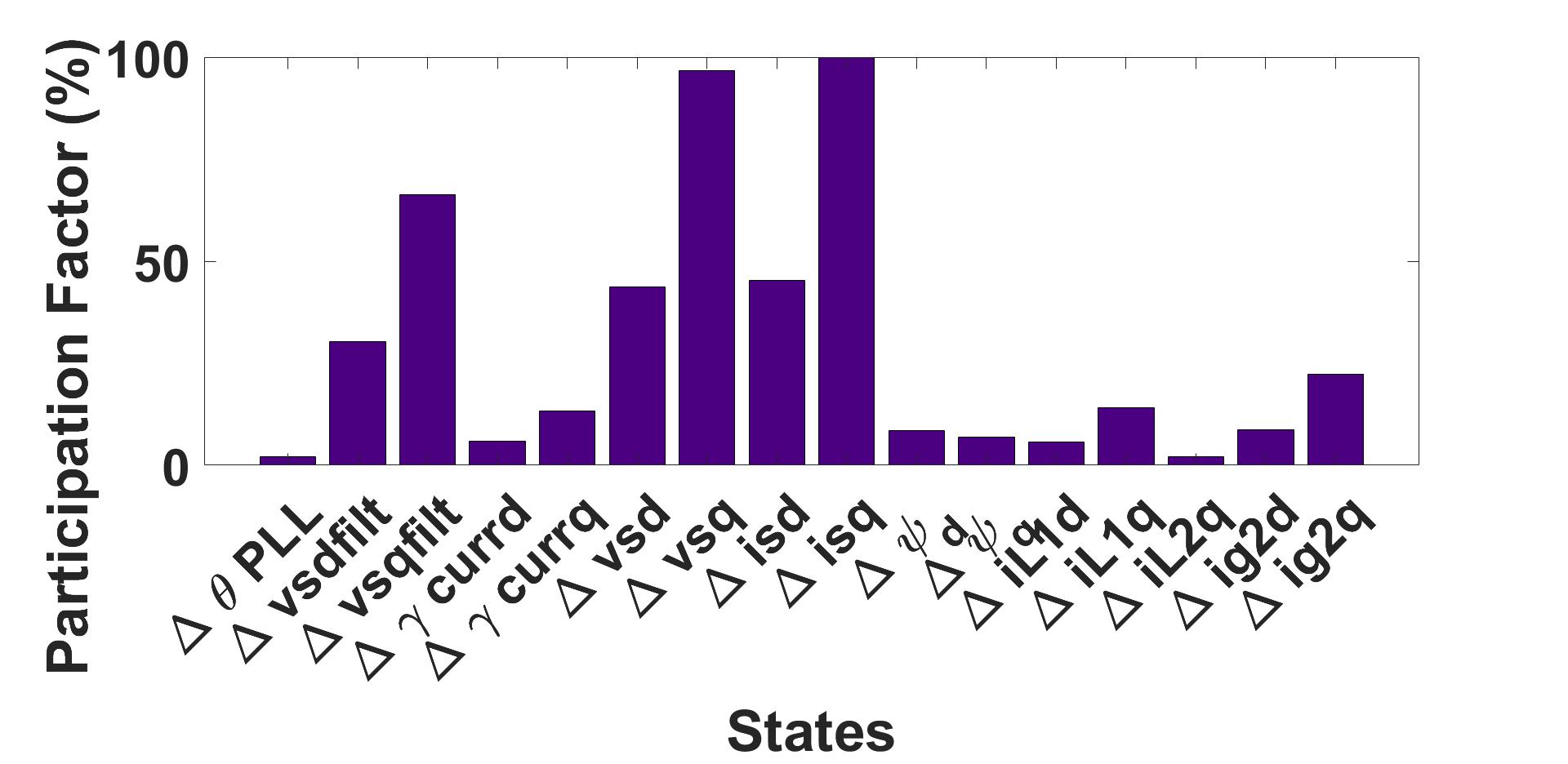}
\caption{Mode 5}
\label{fig:mode 6}  
\end{subfigure}
\caption{\footnotesize{Participation factors}}
\label{image5}
\end{figure*}

To better understand and explain the nature of these modes, the participation factors are calculated. These provide an insight on how each state variable is participating in each mode of oscillation \cite{Kundur2004} \cite{Rogers2015} and are shown in Fig. \ref{image5}.



These modes are all influenced by the inverter current loop states, the inverter filter's states, the PLL, the SM flux linkages and the grid states. Since the inverter, grid and SM states participate in these modes, they can be defined as \textbf{coupling modes}.

$vsdfilt$, $vsqfilt$, $\gamma currd$, and $\gamma currq$ are the states related to the feed-forward filter and the PI current controller respectively.

 Given that almost the same states participate in the different coupling modes but not in the same proportion, a more in-depth examination through sensitivity analyses will be performed in the next section.  
 
\section{Classification of the modes}
As the main objective of this paper is to detect, study and classify in an exhaustive manner the modes of interaction between the SM and the inverter, in addition to participation factors, a sensitivity study against system parameters which might impact the aforementioned modes has been conducted. More precisely, the line lengths, the IBR's control parameters, and the penetration level of the IBR have been varied over the whole range of definition. As the structure of the benchmark has been defined to capture all possible grid situations, the aforementioned parameter variations enabled the identification of general trends applicable to all power system scenarios. In this case the modes have been classified in the category called \textit{structural tendencies} henceforth. The other situations that are non-generic have been classified as \textit{non-structural tendencies}. 
For instance, if a specific mode consistently exhibits increasing damping for a particular grid variation (e.g. for a load decrease or for a decrease of the short-circuit power of the rest of the grid), it indicates a structural tendency. 
\subsection{Structural tendencies}
\underline{Sensitivity to the grid topology}\\

To assess the influence of different grid topologies on the nature and stability of the coupling modes, various tests such as varying $Z_{cc}$, $Z'_3$ and $Z_2$ were conducted for both low-generation ($P_{inv}$ = $P_{sm}$ = 0.2 pu) and high-generation ($P_{inv}$ = $P_{sm}$ = 0.9 pu) operating points with $P_{inv}$ and $P_{sm}$ the injected power by the inverter and the SM respectively. 

First, the nature of the modes in response to varying grid strengths (varying $Z_{cc}$) and varying electrical distances between the IBR and the SM has been studied by examining the participation of the SM's fluxes as the line lengths $l_{cc}$ and $l'_3$ were increased.

As shown in Table \ref{tab:table2}, there is a noticeable increase in the participation of the flux of the SM as the length $l_{cc}$ increases for the low-generation operating point.
This means that when the SM and IBR are connected to a strong grid, the coupling between these two sources is reduced as a result of the predominant impact of the dynamics of the main grid in this situation.

As a result, in the case of a strong grid, the participation of the SM fluxes is reduced in all modes in Table \ref{tab:table1}, and thus the coupling character of the modes is diminished. Consequently, in situations where the coupling is very weak, they may be considered more as local modes of the inverter rather than coupling modes. 

Similar results were observed when the SM and inverter are positioned at a greater distance from each other (i.e. when $Z'_3$ is increased). This result confirms the intuition that the coupling of the modes is stronger when the SM and IBR are closer to each other.
The same results were also confirmed for
the high-generation operating point, thus underscoring a structural tendency.

\begin{table}[h]
\begin{center}
\captionsetup{justification=centering}
\caption{\\ \small{Participation factors of the fluxes in percentage for different values of $l_{cc}$} for $P_{inv}$ = $P_{sm}$ = 0.2 pu}
\addtolength{\tabcolsep}{-2pt} 
\begin{tabular}{ | c | c | c | c | c | c | c | c | c | c | c |} 
  \hline
\textbf{$l_{cc}$}  &\multicolumn{2}{c|}{\textbf{Mode 1}} &\multicolumn{2}{c|}{\textbf{Mode 2}} &\multicolumn{2}{c|}{\textbf{Mode 3}} &\multicolumn{2}{c|}{\textbf{Mode 4}} &\multicolumn{2}{c|}{\textbf{Mode 5}}\\ 
  \hline
    & $\psi d$ & $\psi q$ & $\psi d$ & $\psi q$ & $\psi d$ & $\psi q$ & $\psi d$ & $\psi q$ & $\psi d$ & $\psi q$ \\ 
  \hline
  \textbf{1 Km}  & 0 & 0 & 0 & 0 & 0 & 0 & 0 & 0 & 0 & 0  \\ 
  \hline
  \textbf{5 Km}  & 7.5  & 5.2 & 0 & 0 & 0 & 0 & 7.1 & 7.1 & 7.4 & 6  \\ 
  \hline
  \textbf{10 Km} & 15.5 & 10.5 & 7.7 & 8.7 & 7.4 & 7 & 11.7 & 11.7 & 9.6 & 7.8  \\ 
  \hline
  \textbf{15 Km} & 21.8 & 14.7 & 12.7 & 14.12 & 10.3 & 9.5 & 15.2 & 14.4 & 10 & 8.5 \\ 
  \hline
  \textbf{20 Km} & 26 & 18 & 18.2 & 19.7 & 12.7 & 11.6 & 15.6 & 14.8 & 10 & 9 \\ 
  \hline
		\end{tabular}
\label{tab:table2}
	\end{center}
\end{table}

\begin{figure}[!h]
\centering
 \captionsetup{justification=centering}

\begin{subfigure}[b]{0.24\textwidth}
\centering
\includegraphics[width=\textwidth, height=0.55\textwidth]{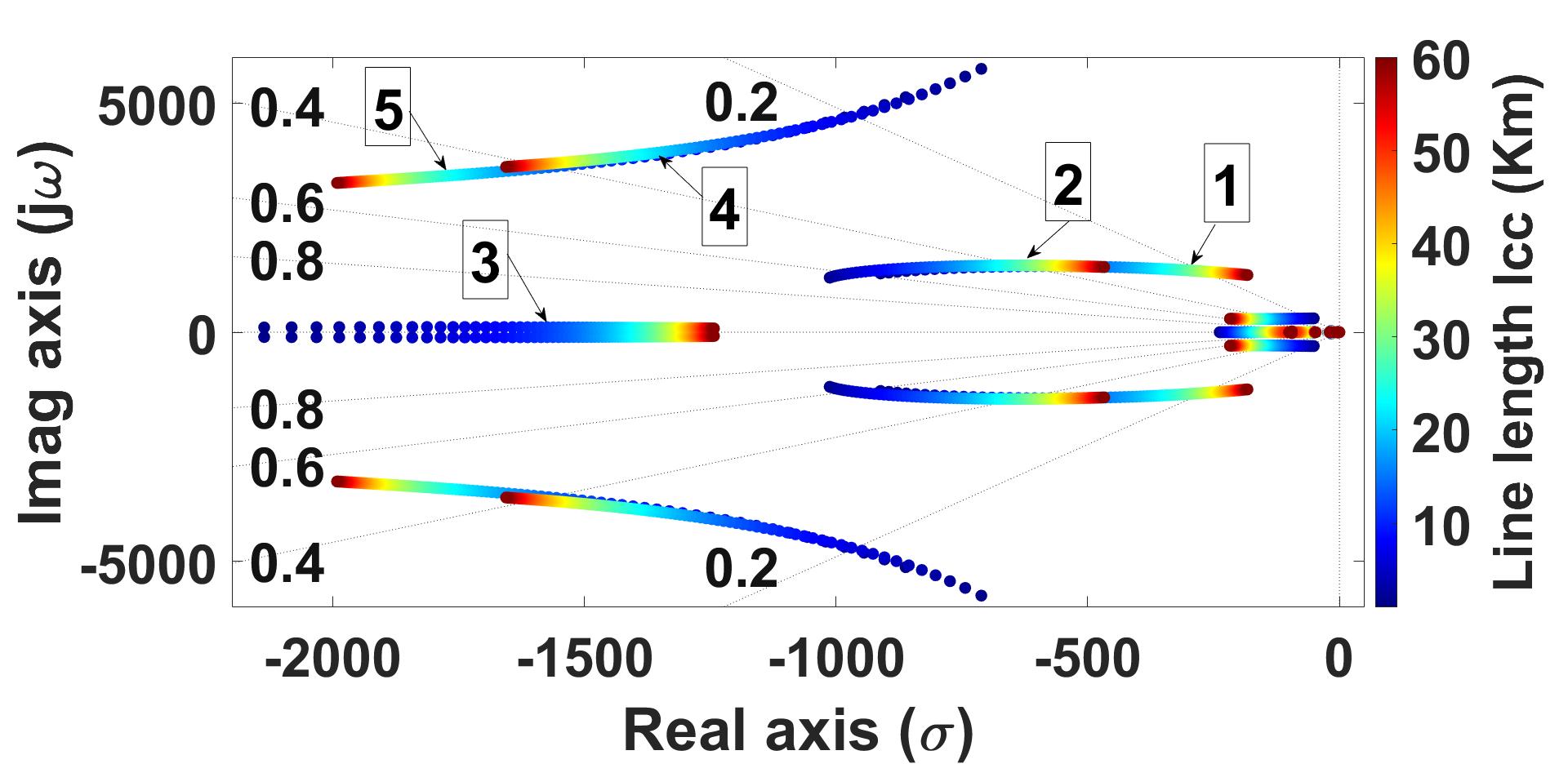}
\caption{$P_{inv}$ = 0.2 pu, $P_{sm}$ = 0.2 pu}
\label{fig:Lcc_op1}  
\end{subfigure}
\begin{subfigure}[b]{0.24\textwidth}
\centering
\includegraphics[width=\textwidth, height=0.55\textwidth]{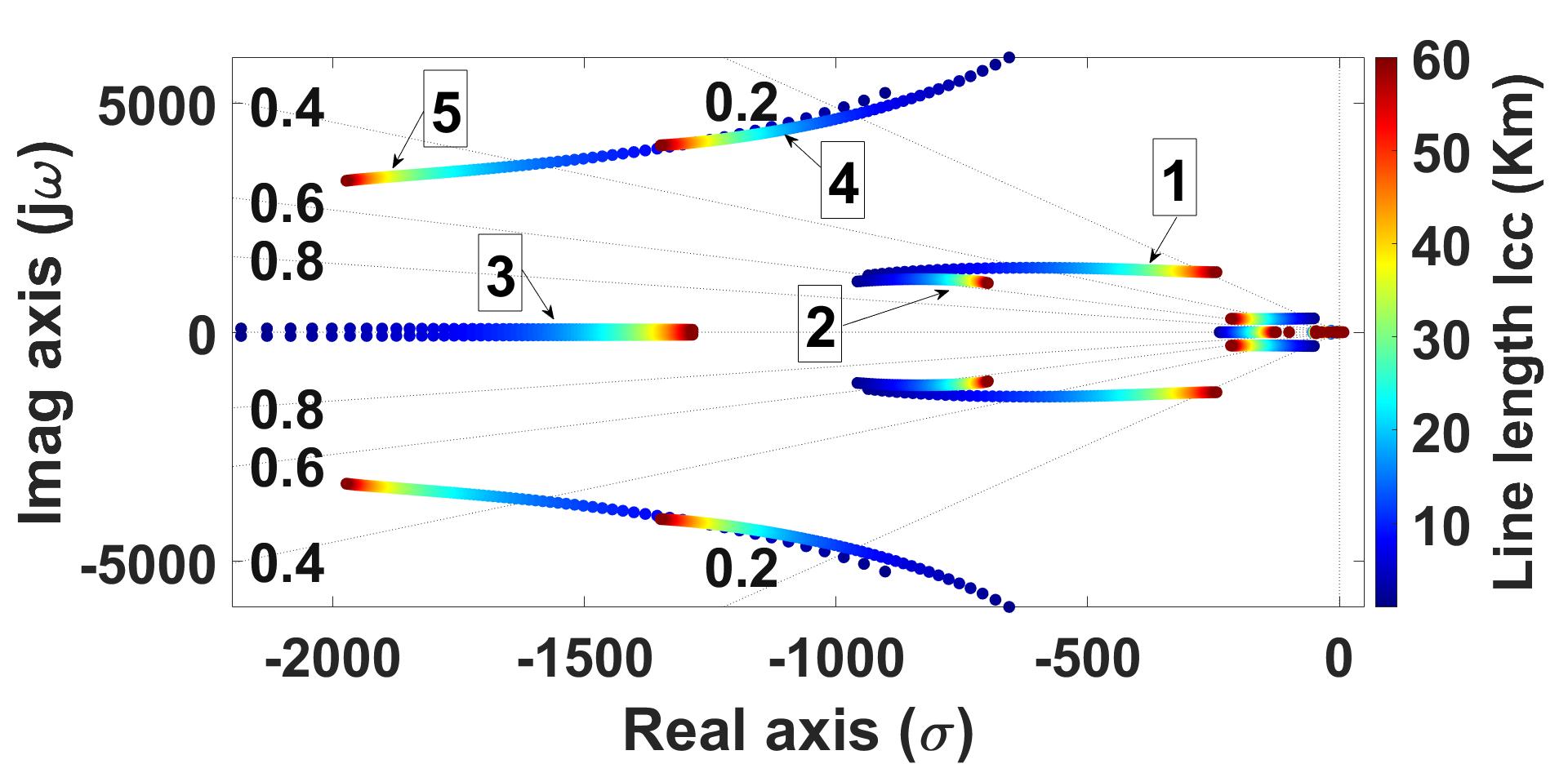}
\caption{$P_{inv}$ = 0.9 pu, $P_{sm}$ = 0.9 pu}
\label{fig:Lcc_op2}  
\end{subfigure}
 \caption{\footnotesize{Eigenvalues sensitivity to $l_{cc}$}}
\label{image6} 
\end{figure}

Next, the variation of the eigenvalues following the variation in $Z_{cc}$ and $Z'_3$ has been studied by varying $l_{cc}$ and $l'_3$. As it is observed in Fig. \ref{image6}, modes 1, 2 and 3 are moving to the right when $l_{cc}$ is increased, hence their damping decreases while the damping of modes 4 and 5 increases (i.e. they become more stable) for the two previously defined operating points. The same results have been observed when varying $l'_3$.

The same tendencies for all the studied modes were observed when increasing either the electrical distance between the inverter and SM ($Z'_3$) or the distance to the main grid ($Z_{cc}$) for both predefined operating points, indicating that their response to variations in $Z'_3$ or $Z_{cc}$ is structural and applicable across various grid scenarios.

\begin{figure}[!h]
\centering
 \captionsetup{justification=centering}
\begin{subfigure}[b]{0.24\textwidth}
\centering
\includegraphics[width=\textwidth]{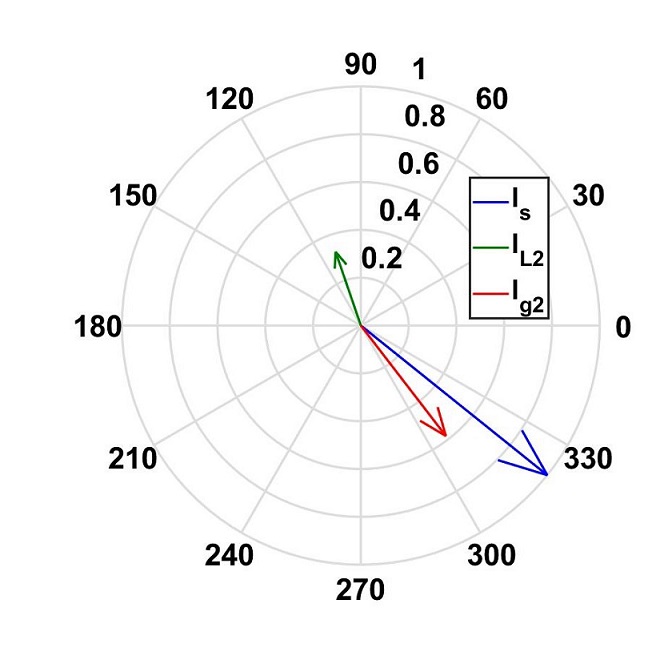}
\caption{Mode 1}
\label{MS:1}  
\end{subfigure}
\begin{subfigure}[b]{0.24\textwidth}
\centering
\includegraphics[width=\textwidth]{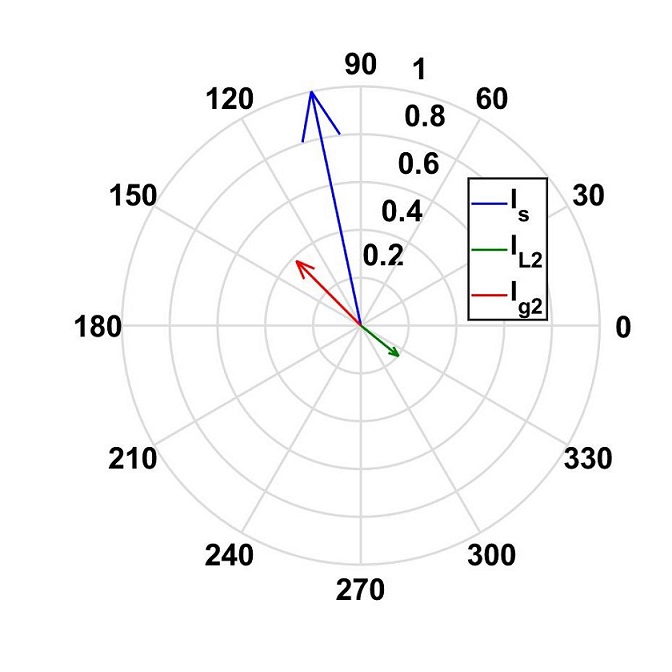}
\caption{Mode 2}
\label{MS:2}  
\end{subfigure}
\\
\begin{subfigure}[b]{0.24\textwidth}
\centering
\includegraphics[width=\textwidth]{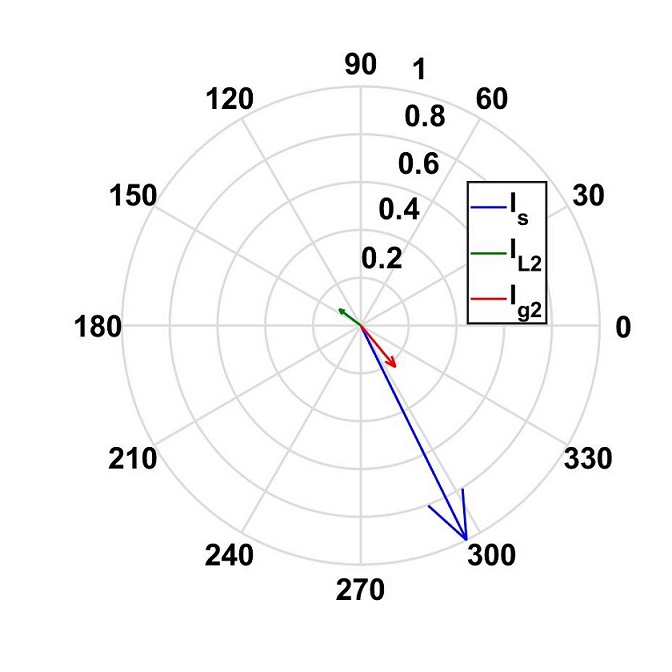}
\caption{Mode 3}
\label{MS:3}  
\end{subfigure}
\begin{subfigure}[b]{0.24\textwidth}
\centering
\includegraphics[width=\textwidth, height = 0.8\textwidth]{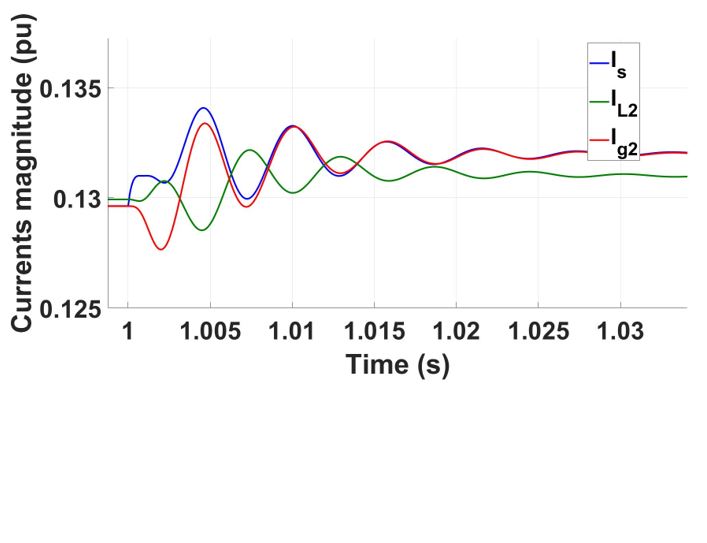}
\caption{Currents magnitudes time domain responses}
\label{current}  
\end{subfigure}
 \caption{\footnotesize{Mode shapes and time-domain simulations for currents magnitude}}
\label{image20} 
\end{figure}

For a deeper understanding of the coupling modes behavior, the extended mode shapes \cite{miao2023} have been calculated according to equation (\ref{eqn 30}). The mode shapes of the  currents magnitudes $I_s$, $I_{L2}$ and $I_{g2}$ flowing into the inverter's filter, the SM and the equivalent grid respectively and corresponding to modes 1, 2 and 3 have been calculated and plotted in Fig. \ref{image20} (\subref{MS:1}-\subref{MS:3}).
The mode shape plots demonstrate that $I_s$ and $I_{g2}$ are oscillating together against $I_{L2}$ in all of the three modes. Similar patterns are observed in nonlinear time-domain plots for mode 1, as depicted in Fig. \ref{image20} (\subref{current}), given its minimal damping and enhanced observability. This consistent trend may offer an explanation for the shared pattern observed in modes 1, 2, and 3 in the previous sensitivity analyses. It suggests that, for these modes, the inverter and the equivalent grid oscillate together against the SM. Also, the same tendencies as in Fig. \ref{image6} were observed when increasing the impedance $Z_2$ between the SM and all the rest of the system including the IBR.

\subsection{Non-structural tendencies}
\underline{Sensitivity to the IBR current controller parameters} \\

\begin{figure}[h]
\centering
 \captionsetup{justification=centering}
\begin{subfigure}[b]{0.24\textwidth}
\centering
\includegraphics[width=\textwidth, height=0.55\textwidth]{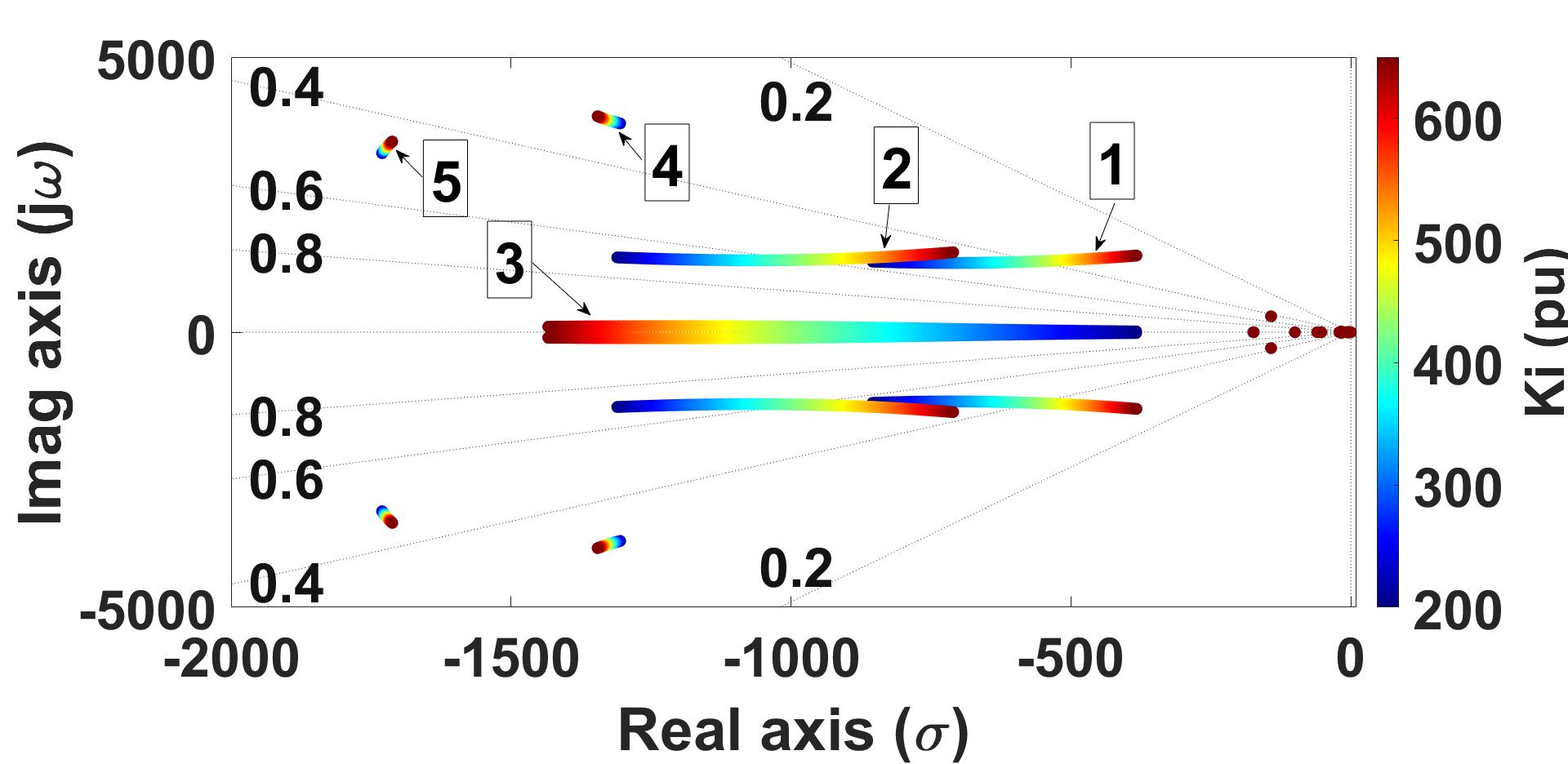}
\caption{$P_{inv}$ = 0.2 pu, $P_{sm}$ = 0.2 pu}
\label{fig:Ki_op1}  
\end{subfigure}
\begin{subfigure}[b]{0.24\textwidth}
\centering
\includegraphics[width=\textwidth, height=0.55\textwidth]{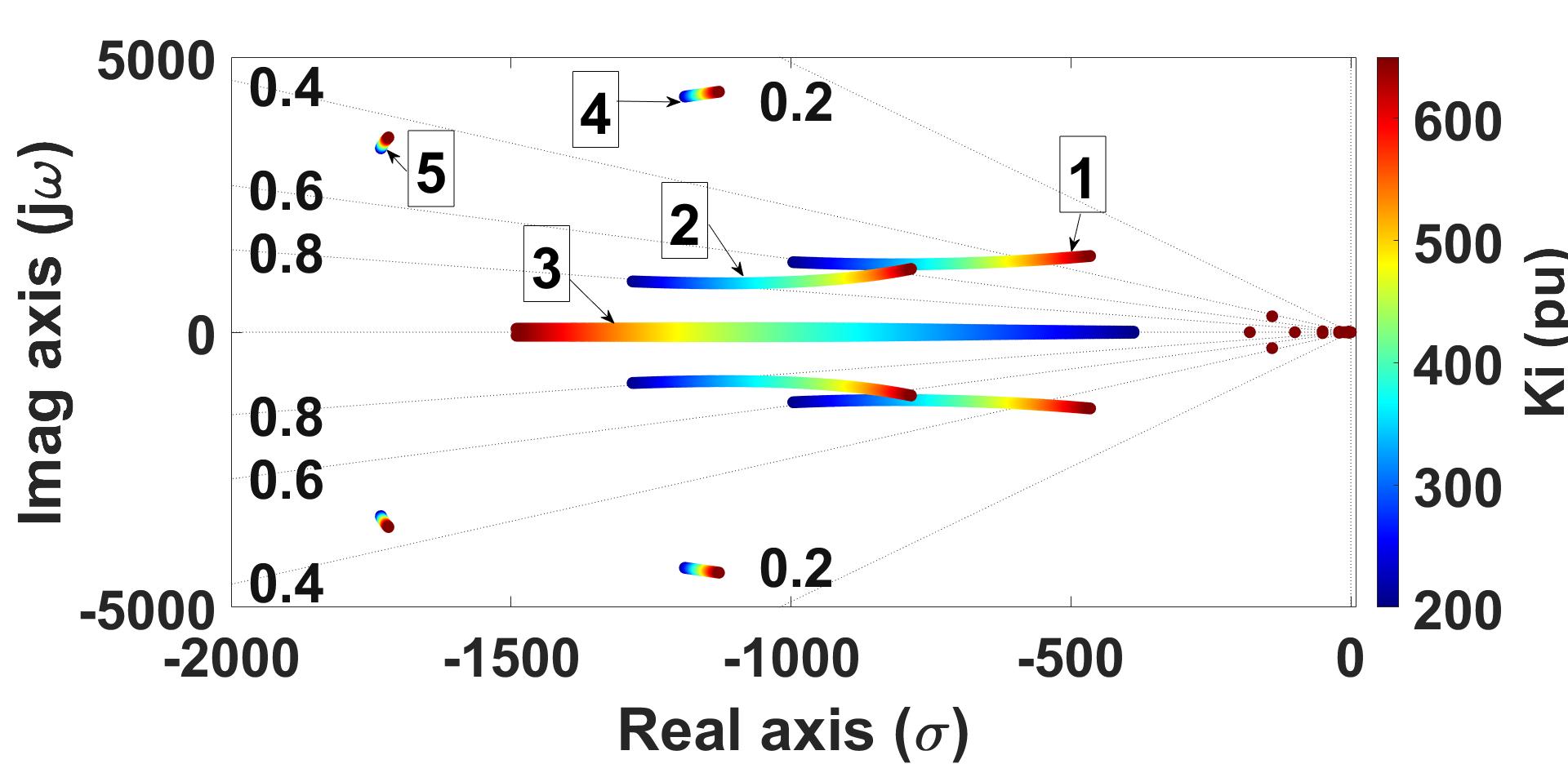}
\caption{$P_{inv}$ = 0.9 pu, $P_{sm}$ = 0.9 pu}
\label{fig:Ki_op2}  
\end{subfigure} 
 \caption{\footnotesize{Eigenvalues sensitivity to the integral gain: $K_i$}}
\label{image8}
\end{figure}

Considering the varying degrees of participation of the states of the inner current loop of the converter in the coupling modes, a sensitivity analysis was performed by varying the parameters of the PI current controller. 
The integral gain was varied from 200 to 300 pu as illustrated in Fig. \ref{image8}.

The results reveal consistent 
behavior among 
modes 1, 2, 3, and 5 across both operating points in response to the variation of $K_i$. However, mode 4 exhibits contrasting 
tendencies for the two operating points, introducing complexity 
into the analysis.

To gain an analytical explanation of these behaviors, the residues have been calculated for each mode. The residues quantify how sensitive each mode is to changes in a specific parameter \cite{Rogers2015}. By examining the angle of the residue, predictions can be made about the direction in which a mode will move when the parameter $K_i$ is increased. If the angle is in Quadrant I or IV, it means that the mode becomes more stable when the parameter increases. Conversely, if it is in Quadrant II or III, the mode's stability decreases when the parameter increases.

As shown in Table \ref{tab:table3}, the angle of the residue for mode 4 is equal to 104\degree \hspace{0.1cm}(Quadrant II) for operating point 1 (op1) and equal to 55\degree \hspace{0.1cm}(Quadrant I) for operating point 2 (op2), explaining the contrasting behavior observed. Thus, mode 4 does not have a structural tendency.
\begin{table}[h]
\begin{center}
\captionsetup{justification=centering}
\caption{\\ \small{Eigenvalues sensitivity with respect to $K_i$ (residus)}}
\addtolength{\tabcolsep}{-1pt} 
\begin{tabular}{ | c | c| c | c |c |c |} 
  \hline
\textbf{} & \textbf{1} & \textbf{2} & \textbf{3} & \textbf{4}& \textbf{5}\\ 
  \hline
  op1&1.25-0.44i&1.16-0.45i&-1.44-0.1i&-0.07+0.29i&0.047+0.44i\\ 
  \hline
  op2&0.6+0.56i&0.42+0.87i&-1.26-0.01i&0.14+0.2i&0.04+0.42i\\ 
  \hline
		\end{tabular}
\label{tab:table3}
	\end{center}
\end{table}

\underline{Sensitivity to the penetration level of the IBR} \\

To assess the effects of gradually replacing SM with IBRs on coupling modes, and consequently, system stability, the ratio of the powers between the inverter and the SM has been varied (and the total volume of generation kept constant).

\begin{figure}[h]
\centering
 \captionsetup{justification=centering}
\begin{subfigure}[b]{0.24\textwidth}
\centering
\includegraphics[width=\textwidth, height=0.55\textwidth]{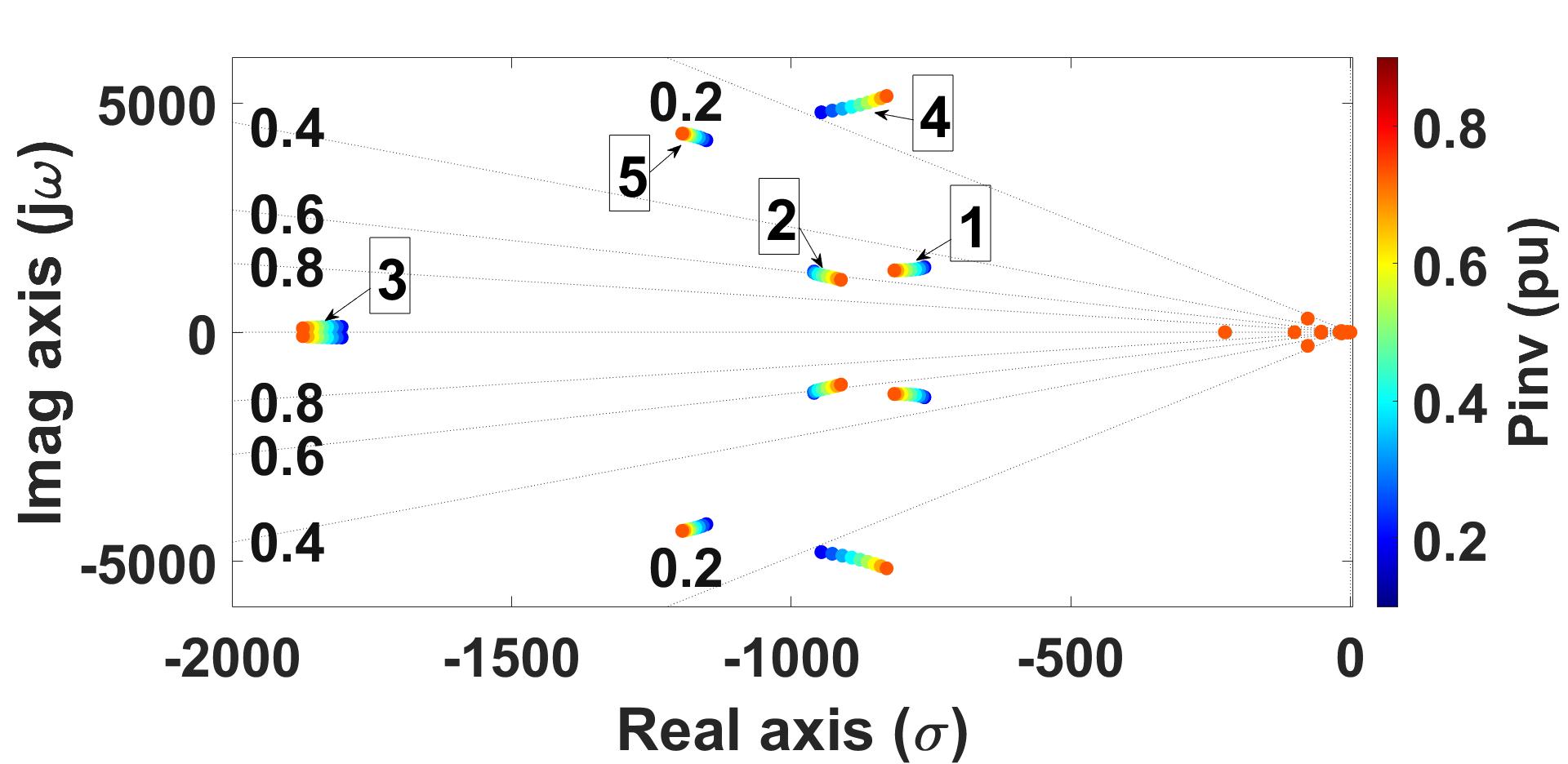}
\caption{$l_{cc}$ = 5 Km}
\label{fig:P_op1}  
\end{subfigure}
\begin{subfigure}[b]{0.24\textwidth}
\centering
\includegraphics[width=\textwidth, height=0.55\textwidth]{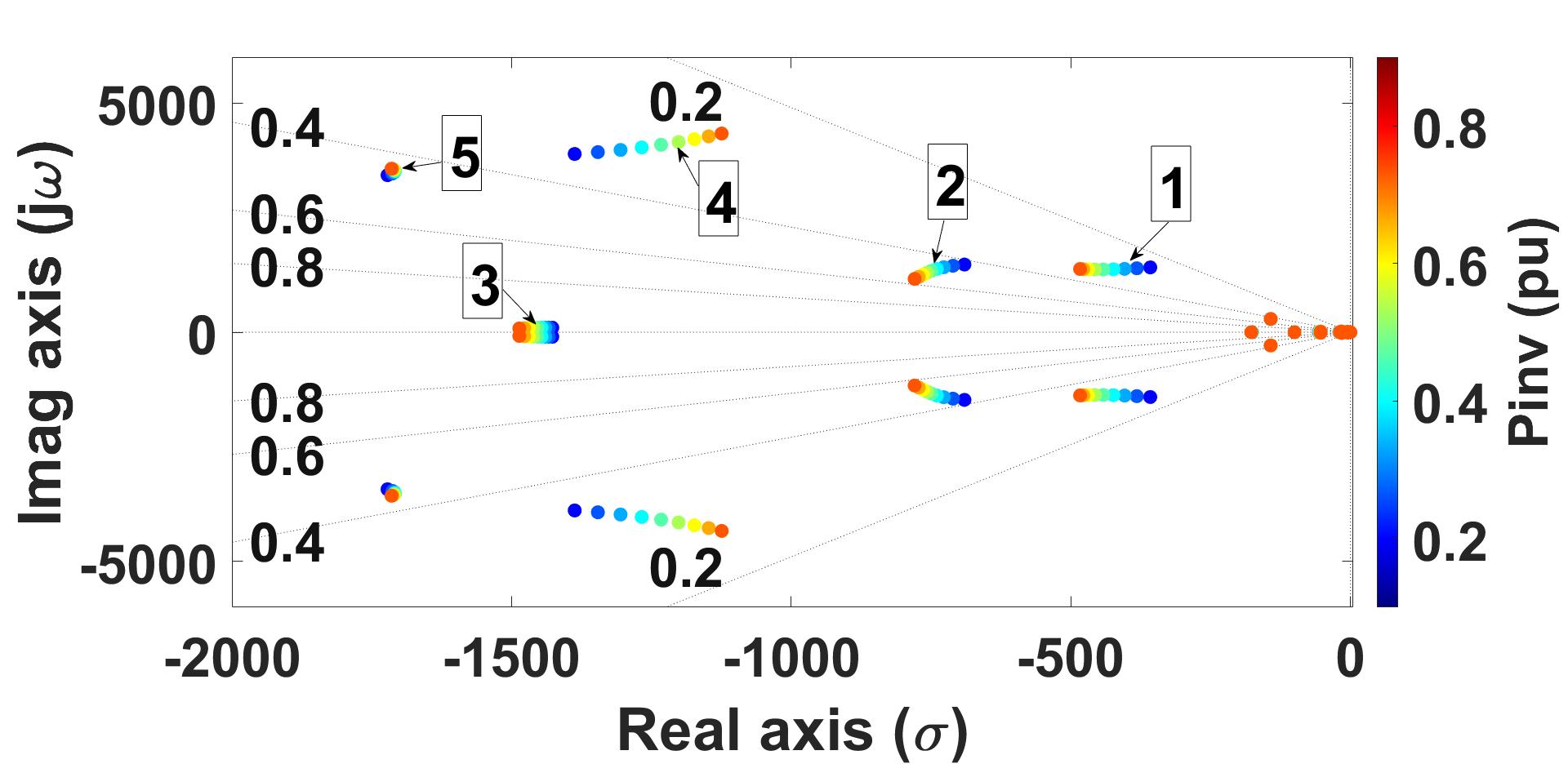}
\caption{$l_{cc}$ = 20 Km}
\label{fig:P_op2}  
\end{subfigure} 
 \caption{\footnotesize{IBR's power injection}}
\label{image9}
\end{figure}

The penetration level of the inverter has been increased by increasing its injected power from 0.1 to 0.9 pu. Two distinct scenarios have been considered: one with a strong grid ($l_{cc}$ = 5 km) and the other one with a weak grid ($l_{cc}$ = 20 km) as shown in Fig. \ref{image9}.

Modes 1, 3 and 4 have the same tendency in both operating points, while modes 2 and 5 exhibit a contrasting behavior in the two scenarios. 
Hence, no generic conclusion can be drawn regarding the displacement of the coupling modes with respect to the IBR's penetration level.

In addition, the impact of the equivalent grid's parameters ($H_e$ and $D_w$) have been studied. However, the analysis showed that these parameters have absolutely no impact on the characteristics of the coupling modes, therefore the results have not been displayed. Moreover, this finding is intuitive given that the coupling modes have high frequencies while the equivalent grid has much slower dynamics due to the swing equation.

The collective findings presented in this section emphasize the complex nature of interaction phenomena in networks containing SMs and IBRs. Nonetheless, modal analysis revealed clear trends and further comprehension of the coupling modes.


\section{Application to the IEEE 39-bus System}
\begin{figure}[h]
	\begin{center}
		\includegraphics[width=8cm,height=7cm]{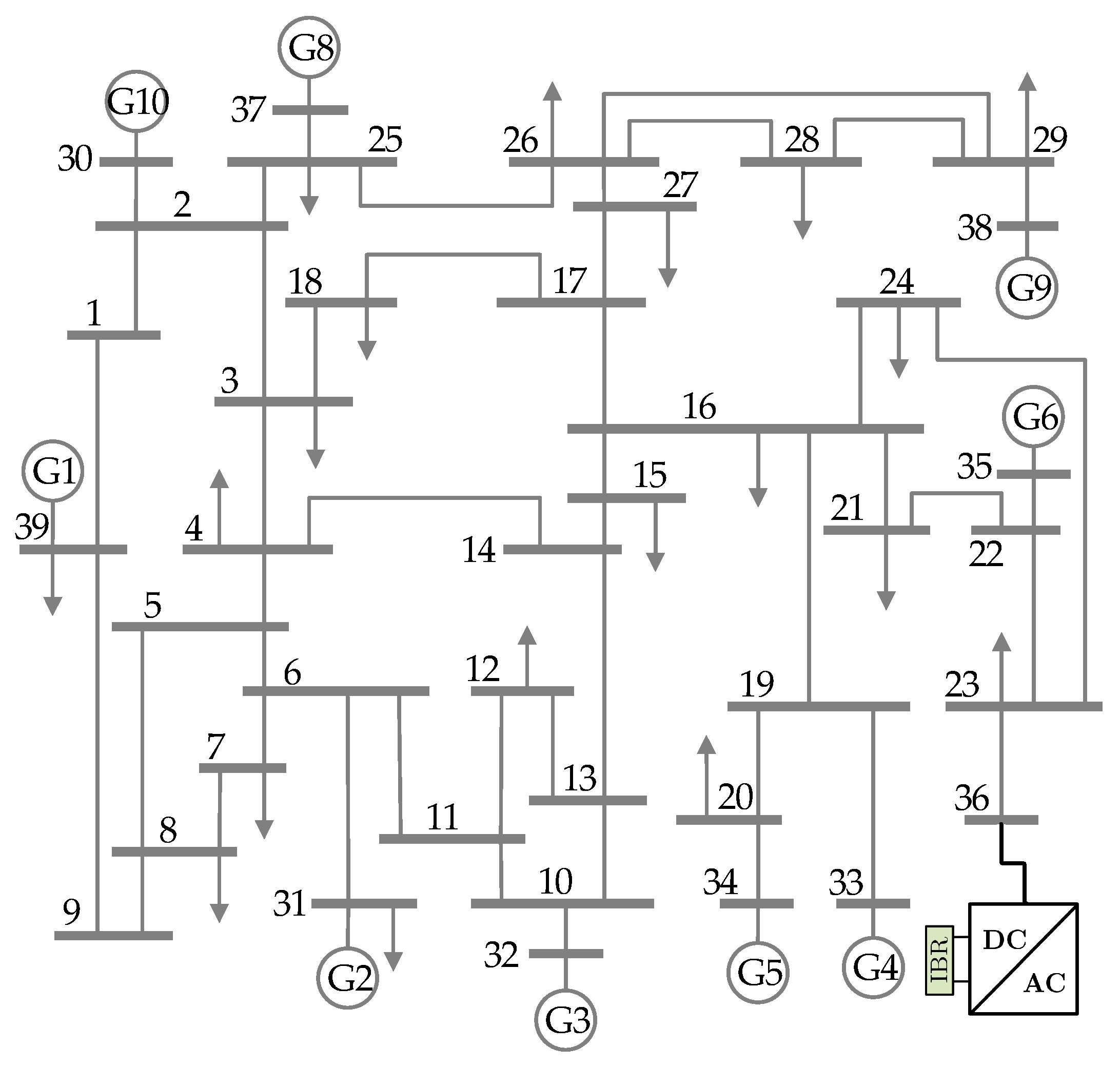}
   \captionsetup{justification=centering}
		\caption{\footnotesize{Modified IEEE 39-bus power system}}
		\label{image10}
	\end{center}
\end{figure}

To demonstrate the benchmark's effectiveness in capturing coupling modes and representing a real, detailed power system, the IEEE 39-bus system, illustrating the New England power system, has been employed as a case study.
Generator 1 at bus 39 shown in Fig. \ref{image10}, is an aggregation of a large number of generators symbolizing the New York power system interconnected to the New England power system. The system's parameters and transmission line data are provided in \cite{Pai1989}. 

\begin{figure}[h]
\centering
 \captionsetup{justification=centering}

\begin{subfigure}[b]{0.24\textwidth}
\centering
\includegraphics[width=\textwidth, height=0.55\textwidth]{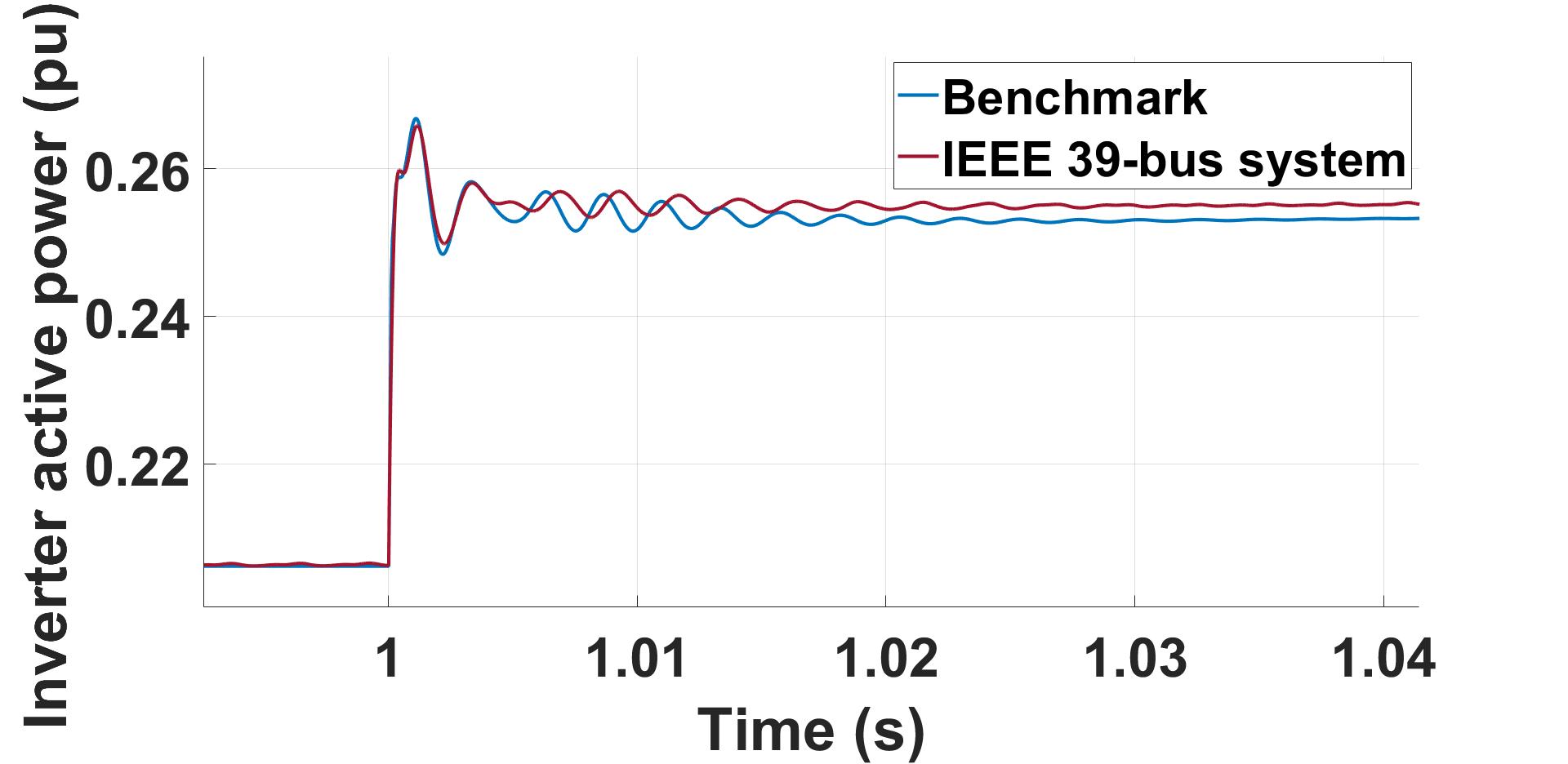}
\caption{}
\label{timedomain}  
\end{subfigure}
\begin{subfigure}[b]{0.24\textwidth}
\centering
\includegraphics[width=\textwidth, height=0.55\textwidth]{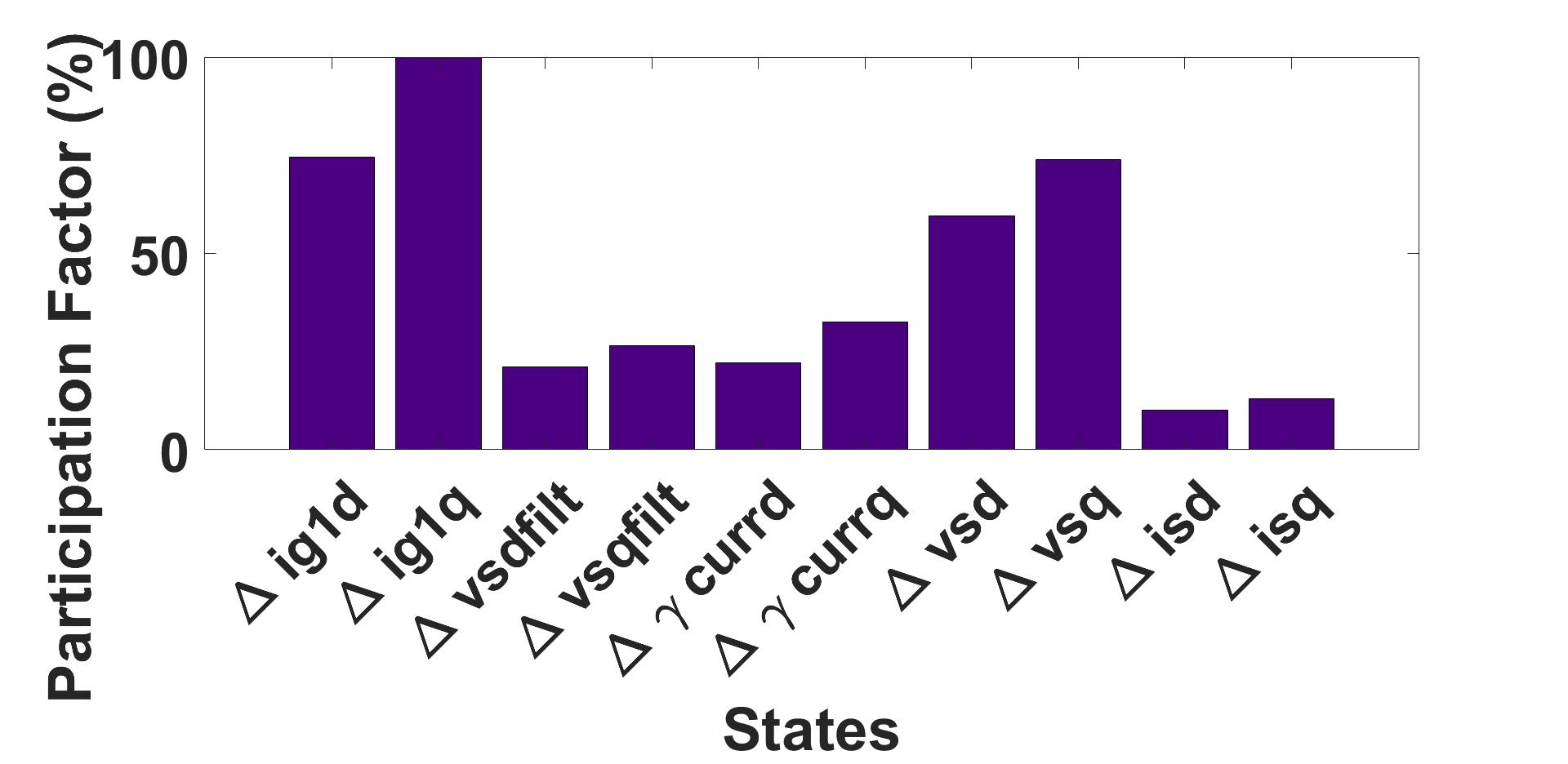}
\caption{}
\label{PF}  
\end{subfigure} 
 \caption{\footnotesize{ (a) Power measured after the inverter's filter. (b) Participation factors of the observed mode }}
\label{image11}
\end{figure}

The system has been modified to reflect a scenario involving a combination of SMs and IBRs. Generator 1 has been eliminated to avoid any bias from an equivalent grid model that could potentially mask some of the detailed coupling interactions within the grid. Generator G7 on bus 36 has been replaced by an IBR controlled in grid-following mode and generator G6 on bus 35 has been replaced by the same SM employed in the benchmark in section II. \\
In this case study, the aim is to examine the interactions between the IBR and the generator G6 at node 35. This will be accomplished by employing the benchmark approach, where the remainder of the network will be aggregated in the equivalent network model as described in section II.D. To reproduce this situation using the benchmark's configuration in Fig. \ref{image1} (\subref{fig:triangle}), first, the equivalent line impedance parameters have been estimated using the following method in simulation: 
\begin{itemize}
  \item
The equivalent impedance $Z_2'$ between the SM G6 and the rest of the system is determined by applying a short circuit near the SM and measuring the current in line 21-22 of Fig. \ref{image10}.
\item Having the short circuit current $I_{cc}$, the impedance is calculated as $Z= \frac{U}{I_{cc}\sqrt3}$ with U = 500 kV.
\item The same procedure is applied for computing $Z_1'$, but this time the short circuit is applied in proximity to the inverter, and the current flowing through line 23-24 is used in the calculation.
\item The inertia of the equivalent grid $H_t$ is calculated as $H_{t} = \frac{\sum H_iS_i}{S_t}$, with $H_i$, $S_i$ the inertia and rated power of each generator respectively, and $S_t$ the total rated power \cite{Sauer2007}.
\item $D_u$ represents the static power-frequency droop relationship of the equivalent grid, and was therefore tuned to match the overall primary frequency control gain of the modified IEEE 39-bus system. 
\end{itemize}
The adapted benchmark was validated against the modified IEEE 39-bus power system using EMTP software. For this purpose, a step of amplitude 0.05 pu has been applied to the inverter's active power setpoint. The dynamic responses to this step using the adapted benchmark and the modified IEEE 39-bus system are illustrated in Fig. \ref{image11} (\subref{timedomain}). In both simulation results, the same oscillatory mode with a frequency close to 415 Hz and a damping of about 5 \% can be observed. This mode observed in the nonlinear time-domain simulations has been detected in the modal analysis and corresponds the mode 1 studied in the previous sections. Its participation factors are illustrated in  Fig. \ref{image11} (\subref{PF}). Only states related to the inverter are participating in this mode and there is no participation of the SM fluxes, hence there is no coupling and the mode can be considered to be local to the inverter.

The impedances in this case are relatively high, indicating a considerable length of the three lines of Fig. \ref{image1} (\subref{fig:triangle}). This condition implies that both sources are not only far from the rest of the grid, but also distant one from each other. These findings align with the low damping of the mode and the weak coupling, consistent with the results of the sensitivity analysis of the previous section. Therefore, improvement in damping can be achieved by adjusting the controller parameters or by reducing the line impedances, such as through line reinforcement.


\section{Conclusion}

In this study, an exhaustive classification and quantification of the possible coupling modes between the SMs and IBRs in power systems has been conducted. For that, an original minimal size benchmark representative of all possible situations of a power system relevant for such coupling modes has been introduced. It has been concluded that the aforementioned coupling modes may exist at various frequencies between dozens to hundreds of hertz. Analysis is more difficult than for the classical inter-area modes as oscillation patterns may be more complex. Indeed, an extension of the classic mode shape concept can be used to put into evidence phase and anti-phase swings for several state variables - and not only for the generators speed as in the case of electromechanical inter-area modes - for a single given mode. Also, several modes may be attached to almost similar kind of coupling. To gain deeper insights into such coupling modes, cross-investigations of combinations of basic modal analysis, participation factors, sensitivity analyses and nonlinear simulations are necessary. The study revealed two types of behavior:

\textit{Structural tendencies}: a mode or a set of modes exhibited consistent mechanisms of variations in response to changes in grid strength, electrical distance between SM and IBR and for several critical operating points.

\textit{Non-structural tendencies}: modes with non-structural behavior displayed different evolutions under different grid conditions (for different operating points). For instance, mode 4 exhibited contrasting tendencies in response to variations in IBR's PI controller parameters. This means that no general rule can be formulated and highlights the need to consider specific scenarios when studying the sensitivity of the modes to the current control parameters. 

The impact of replacing SMs with IBRs on coupling modes was also explored for different grid strengths. Notably, some modes (modes 2 and 5) exhibited non-generic tendencies, emphasizing the importance of scenario-specific considerations when assessing IBR penetration levels.

Particularly, and as an improvement compared to previous investigations in the literature, the proposed benchmark and methodology facilitated the consideration of the impact of the rest of the grid on the coupling modes. When the rest of the grid is weak or when the SM and the inverter are in close proximity, a strong coupling is evident, as indicated by high participation factors. This insight is valuable because it can help network operators and utilities determine the relevant type of stability studies required based on network topology. When the topology exhibits weak coupling (quasi-local mode), conventional studies like the IBR vs an infinite bus may be sufficient to assess stability and control risks. However, in scenarios with strong coupling, a comprehensive model that accounts for both the SM and the equivalent power system is essential for stability characterization. In these situations, a standard infinite bus study would not accurately represent the system's behavior.

\begin{figure}[htbp]
	\begin{center}
	\includegraphics[width=8.5cm]{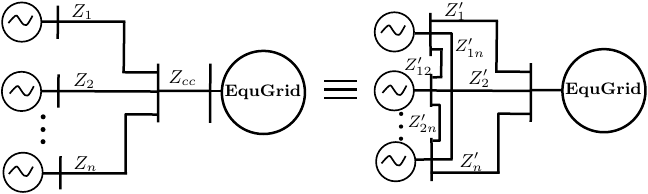}
   \captionsetup{justification=centering}
		\caption{\footnotesize{Adapted benchmark for multiple components}}
		\label{image12}
	\end{center}
\end{figure}

Moreover, the proposed framework allows one to study the general situation with several SMs and IBRs for which the structure of the benchmark must be adapted as in Fig. \ref{image12}, based on the specific research requirements \cite{Marinescu2021}. 

In addition, as it covers all grid situations, the proposed benchmark can also be used to investigate coupling modes of any given power system in a minimal order model. Indeed, any power system can be reduced to the structure of Fig. 1 or Fig. 12, as shown for the modified IEEE 39-bus. On such minimal structure, the mechanism of the coupling modes can be emphasized rapidly and this can provide an important basis for investigations with full detailed models. 

The grid-forming control mode and the influence of the tuning of the PLL on the coupling modes will be studied in future works with this framework.

\printbibliography
\section*{Appendix A}
\begin{table}[H]
\begin{center}
\captionsetup{justification=centering}
\caption{\\ \small{System Parameters}}
\addtolength{\tabcolsep}{2pt}
\begin{tabular}{l l l}
\multicolumn{3}{c}{\textbf{Inverter's Filter Parameters}} \\
			\hline
   \hline
{Filter resistance} & {($R_{f}$, p.u.)} & {0.015}\\
 \hline
{Filter inductance} & {($L_{f}$, p.u.)} & {0.1}\\
 \hline
{Filter capacitance} & {($C_{f}$, p.u.)} & {0.11}\\
\hline\hline
\multicolumn{3}{c}{\textbf{Inverter's Control Parameters}} \\
			\hline
   \hline
\multicolumn{3}{c}{\textbf{Power Control}} \\
			\hline
{Droop gain} & {($R_{p}$, \%)} & {4}\\
 \hline
 {Frequency estimation filter time constant} & {($T_{w}$, ms)} & {100}\\
 \hline
{Power measurement filter time constant} & {($\tau_{p}$, ms)} & {0}\\
 \hline
 \multicolumn{3}{c}{\textbf{Current Control}} \\
\hline
 {Feed-Forward filter time constant} & {($\tau_{f}$, ms)} & {0.3}\\
 \hline
{Integral gain} & {($K_{i}$, p.u.)} & {650}\\
 \hline
 {Proportional gain} & {($K_{p}$, p.u.)} & {0.62}\\
 \hline
 \multicolumn{3}{c}{\textbf{Phase Locked Loop}} \\
			\hline
{Integral gain} & {(${K_{i}}_{pll}$, p.u.)} & {9.82}\\
 \hline
{Proportional gain} & {(${K_{p}}_{pll}$, p.u.)} & {0.3183}\\
 \hline  \hline
\multicolumn{3}{c}{\textbf{Equivalent Grid Parameters}} \\
			\hline
   \hline
{Damping coefficient} & {($D_{u}$, p.u.)} & {0}\\
 \hline
{Inertia constant} & {($H_{e}$, s)} & {1.438}\\
 \hline
\multicolumn{3}{c}{\textbf{Synchronous Machine Parameters}} \\
			\hline
   \hline
{Rated frequency} & {($f_b$, Hz)} & {50}\\
 \hline
{Inertia constant} & {($H$, s)} & {1.438}\\
 \hline
{Damping coefficient} & {($K_d$, p.u.)} & {0}\\
 \hline
 \multicolumn{3}{c}{\textbf{AVR Parameters}} \\
			\hline
   \hline
{Transducer time constant} & {($T_e$, ms)} & {50}\\
 \hline
 {Exciter gain} & {($K_e$, p.u.)} & {500}\\
 \hline
\multicolumn{3}{c}{\textbf{Frequency Control Parameters}} \\
			\hline
   \hline
{Droop gain} & {($R_{g}$, \%)} & {2}\\
 \hline
{Governor time constant} & {($T_{g}$, ms)} & {500}\\
 \hline
 {Turbine reheat time constant} & {($T_{r}$, ms)} & {10 000}\\
 \hline
  {Turbine power fraction factor} & {($F_{h}$, -)} & {0.1}\\
 \hline
 \multicolumn{3}{c}{\textbf{Lines impedance}} \\
			\hline
   \hline
{$Z_1$ = $Z_2$ = $Z_{cc}$} & {($\Omega/Km$)} & {0.05}\\
 \hline
 {Ratio $X/R$}& {} & {1.04}\\
 \hline
		\end{tabular}
\label{tab:table4}
	\end{center}
\end{table}

\end{document}